\documentclass[12pt,a4paper]{article}
\usepackage[english]{babel}
\usepackage{graphicx}

\newcommand{\alt}{\mathbin{\lower 3pt\hbox
   {$\rlap{\raise 5pt\hbox{$\char'074$}}\mathchar"7218$}}}
\newcommand{\agt}{\mathbin{\lower 3pt\hbox
   {$\rlap{\raise 5pt\hbox{$\char'076$}}\mathchar"7218$}}}

\begin{document}
\setcounter{footnote}{0}
\setcounter{equation}{0}
\setcounter{figure}{0}
\setcounter{table}{0}
\vspace*{5mm}

\begin{center}
{\large\bf Finite-size scaling \\
from self-consistent theory of localization }

\vspace{4mm}
I. M. Suslov \\
P.L.Kapitza Institute for Physical Problems,
\\ 119334 Moscow, Russia \\
E-mail: suslov@kapitza.ras.ru
\\
\vspace{5mm}
\end{center}

\begin{center}
\begin{minipage}{135mm}
{\large\bf Abstract } \\
Accepting validity of self-consistent theory of
localization by Vollhardt and W${\rm {\ddot o}}$lfle,
we derive the finite-size scaling procedure used for
studies of the critical behavior in  $d$-dimensional case
and based on the consideration of auxiliary quasi-1D systems.
The obtained scaling functions for $d=2$ and $d=3$ are in good
agreement with numerical results: it signifies
the absence of essential contradictions with
the Vollhardt and W${\rm {\ddot o}}$lfle theory
on the level of  raw data. The results  $\nu=1.3\,-\,1.6$,
usually obtained at  $d=3$ for the critical exponent  $\nu$
of the correlation length, are explained by the fact that
dependence  $L+L_0$ with $L_0>0$ ($L$ is the transversal
size of the system) is interpreted as $L^{1/\nu}$ with $\nu>1$.
For dimensions $d\ge 4$,  the modified scaling relations are
derived; it demonstrates incorrectness of the conventional
treatment of data for $d=4$ and  $d=5$, but establishes  the
constructive procedure for such a treatment.
Consequences for other variants of finite-size scaling are
discussed.

\end{minipage}
\end{center} \vspace{5mm}

\begin{center}
{\bf 1. INTRODUCTION}
\end{center}

The contemporary
situation in investigation of the
Anderson localization is characterized by the fact
that the results of numerical modelling  (see a review article
\cite{2}) contradict all other information on the critical
behavior \cite{2,21,101}. Such situation is unacceptable,
since undermines a belief in analytical theory.

The critical behavior of conductivity $\sigma$  and the
correlation length $\xi$
$$
\sigma \propto \tau^s\,,\qquad \xi \propto |\tau|^{-\nu}
\eqno(1)
$$
($\tau$ is a distance to the transition point)
obtained from the  self-consistent theory of localization
by Vollhardt and W${\rm {\ddot o}}$lfle  \cite{22,23},
has a form
$$
\nu=\left \{
\begin{array}{cc} 1/(d-2)\,,&\quad 2<d<4
\\ 1/2\,, & \quad d>4
\end{array}
\right.\,, \qquad\quad s=1\,, \quad  2<d<\infty  \,,
\eqno(2)
$$
($d$ is the dimension of space), and in fact summarizes
all known results.  Indeed, the formula (2):

 (a) distinguishes values  $d_{c1} = 2$ and $d_{c2} = 4$ as the
 lower  and upper critical dimensions, which are known from
 independent  arguments (see  \cite{21,25} for details);

(b) agrees with theory for  $d=2+\epsilon$ \cite{29}
$$
\nu = \frac{1}{\epsilon} + 0 \cdot \epsilon^0 + 0 \cdot \epsilon^1 + O
(\epsilon^2) \,;
\eqno(3)
$$

(c) satisfies the Wegner scaling relation  $s=(d - 2) \nu$
\cite{1} for $d<d_{c2}$;

(d) gives independent of  $d$ critical exponents for $d > d_{c2}$,
as
it is typical
for the mean-field theory;

(e) agrees with the results  $\nu = 1/2$  \cite{31,32}
and $s = 1$ \cite{33} for $d = \infty$;

(f) agrees with the experimental results $s\approx 1$,
$\nu\approx 1$ for  $d=3$, obtained by the measurement of
conductivity and dielectric susceptibility
\cite{34,35}.\,\footnote{\,The paper \cite{35} is especially
interesting, since the experiment is made for
the nondegenerate electron gas and the influence of
interaction can be controlled explicitly.}

It is clear that the  Vollhardt and W${\rm {\ddot o}}$lfle
theory gives at least a very successful approximation, satisfying
all general principles and reproducing all known results.
More than that, a suspicion arises that the result  (2)
is exact \cite{36}\,\footnote{\,According to Wegner \cite{30}, the term
 $O(\epsilon^2)$ in (3) is finite  and large negative.
However, this result was derived for the
zero-component $\sigma$-model, whose correspondence with the initial disordered system
is approximate and valid for small $\epsilon$; so a difference can arise in a
certain order in $\epsilon$. }.
 This
conjecture is supported by the paper  \cite{37}, where Eq.\,2 is
derived without model approximations on the basis of symmetry
analysis.

As for numerical results \cite{8}--\cite{20a},
they can be summarized by the empirical formula
$\nu \approx 0.8/(d-2) +0.5$ \cite{17}, which has the
evident fundamental defects. Recent developments
make a situation even worse giving for $d=3$
values $\nu=1.54\pm 0.08$ \cite{16},
$\nu=1.45\pm 0.08$ \cite{18},
$\nu=1.40\pm 0.15$ \cite{18a}, $\nu=1.57\pm 0.02$ \cite{19}, etc.

In our opinion, it means the existence of serious defects in
the conventional numerical algorithms.
It is not reasonable to call in question the raw
data, which are obtained independently by many groups; but it is
possible to doubt the algorithms themselves, which are not
based on any serious theory.
In particular, there is a possibility
of rough violation of scaling \cite{102}, or
existence of the large characteristic length scale
\cite{100,101}.

In the present paper,  the following approach is accepted. We
suppose that the  Vollhardt and W${\rm {\ddot o}}$lfle theory
(Sec.\,2)
is correct (there are real grounds for such assumption \cite{37})
and derive the quantities which are immediately "measured"
in the numerical experiments. Then comparison can be made on the
level of the raw data, avoiding  the suspicious treatment
procedure.

We restrict the discussion by the popular variant of
finite-size scaling based on consideration of auxiliary
quasi-1D systems \cite{7}. Thus, instead of the infinite 3D
system one consider the system of size $L\times L \times L_z$,
where  $L_z \to\infty$. Such system is topologically
one-dimensional and does not possess the long-range order:
so the corresponding correlation length $\xi_{1D}$ is
finite. If $\xi_{1D}$ can be calculated, then its dependence
at $L\to\infty$ allows to registrate  phase transitions in
the initial $3D$  system: it appears, that
$\xi_{1D}/L\to\infty$   in the phase with
long-range order and $\xi_{1D}/L\to 0$ in the phase with
 short-range correlations \cite{7,102}.
In the numerical studies,
the following
scaling relation is usually postulated
$$
\frac{\xi_{1D}}{L} = F\left(\frac{L}{\xi} \right) \,.
\eqno(4)
$$
It is based on the assumption that the correlation length  $\xi$
of the considered $d$-dimensional system is
the only essential length scale, so  $L$ enters only in the
combination  $L/\xi$. If this relation is valid, then
the quantity $\xi_{1D}/L$ depends on $L$ in accordance with
Fig.\,1:
\begin{figure}
\centerline{\includegraphics[width=5.1 in]{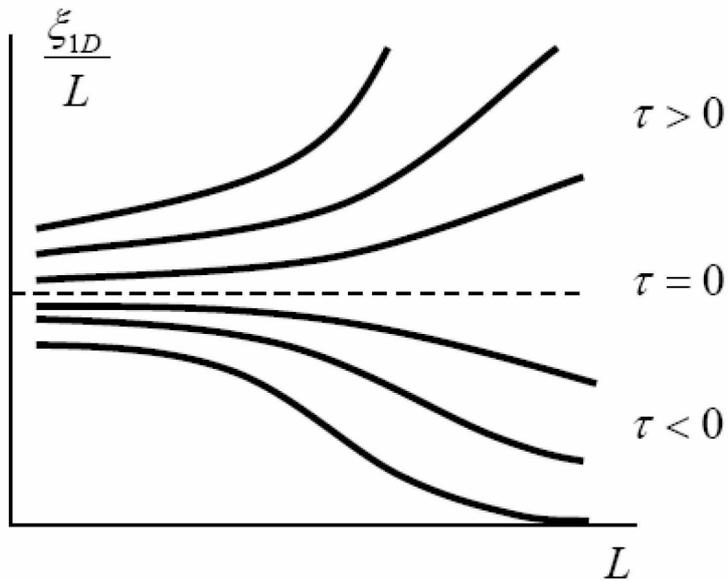}}
\caption{Dependence of the scaling parameter
$\xi_{1D}/L$  on the transversal size $L$ of the system.}
\label{fig1}
\end{figure}
it remains constant at the critical point,
while all curves for $\tau>0$ (and correspondingly $\tau<0$) can
be reduced to one universal curve by the scale transformation.
If two curves for $\tau=\tau_1$ and $\tau=\tau_2$ are calculated,
then the scale transformation allows to determine the ratio of two
correlation lengths.
%
Taking succession $\tau_1$,
$\tau_2,\,\ldots$, one can determine $\xi(\tau_i)$
apart from numerical factor and to investigate its  critical
behavior.

We demonstrate below, that the scaling relation  (4)  is
indeed valid in the limit of large  $\xi$ and $L$ for space
dimensions  $d <4$, while calculation of the scaling function
$F$  for $d=2$ and $d=3$ shows a  good
agreement with numerical results (Sec.\,3). It signifies that
 the Vollhardt and W${\rm {\ddot o}}$lfle theory is confirmed
 on the level of raw data.  Section 4 clarifies why  values
of the exponent $\nu$ in the numerical experiments  for $d=3$
are always
greater than unity: in the vicinity of the critical point the
scaling parameter  $\xi_{1D}/L$ behaves as  $\tau(L+L_0)$ with
$L_0>0$, which is conventionally interpreted as  $\tau L^{1/\nu}$
with $\nu>1$.

For higher dimensions, the scaling relation (4) cannot be correct,
and it can be stated on the level of a theorem. The problem of
the Anderson transition can be exactly reduced to the $\phi^4$
field theory \cite{25,26,27,28}, which is
non-renormalizable for $d>4$  \cite{200,201}. Therefore, the
ultraviolet cut-off (i.\,e. atomic scale)  cannot be excluded from
results, and  $\xi$ is certainly not the only relevant
length scale. However, it is possible to derive the modified
scaling relations
$$
y=F(x)
\eqno(5)
$$
with
$$
y=\frac{\xi_{1D}}{L} \left(\frac{a}{L} \right)^{(d-4)/3}\,,\qquad
x=\frac{\xi}{L} \left(\frac{a}{L} \right)^{(d-4)/3}\,,\qquad
d>4
\eqno(6)
$$
and
$$
y=\frac{\xi_{1D}}{ L} \left[\ln(L/a)\right]^{-1/3}
\,,\qquad
x=\frac{\xi}{L} \,
  \frac{\left[\ln(L/a)\right]^{1/6}}
  { \left[\ln(\xi/a)\right]^{1/2}  } \,,\qquad d=4,
\eqno(7)
$$
demonstrating  incorrectness of the conventional treatment of
data for $d=4$ and  $d=5$ \cite{2}, but establishing the
constructive
procedure for such a treatment (Sec.\,5).
The modified scaling (5) with
$$
y=\frac{\xi_{1D}}{ L}
\left[\frac{\epsilon}{1-(L/a)^{-\epsilon}}\right]^{1/3}
\,,\qquad
x= \frac{\xi}{L}  \,
  \frac{\left[1-(L/a)^{-\epsilon} \right]^{1/6}}
  { \left[(\xi/a)^{\epsilon}-1\right]^{1/2}}
  \,\frac{\epsilon^{1/3}}{(L/a)^{-\epsilon/2} }
  \,, \qquad d=4-\epsilon
\eqno(8)
$$
can be derived also for $d=4-\epsilon$ dimensions.
It can be used for alternative treatment at $d=3$,
in order to estimate the systematic errors related with
possible existence of the large length scale.
Finally, in Sec.\,6 we discuss some consequences of the present
analysis  for other variants of finite-size
scaling.

\vspace{2mm}

\begin{center}
{\bf 2. VOLLHARDT AND W${\rm {\ddot O}}$LFLE THEORY
}
\end{center}

The Vollhardt and W${\rm {\ddot o}}$lfle theory is based
on existence of the diffusion pole in the irreducible
four-leg vertex  $U_{{\bf k} {\bf k}^\prime} ({\bf q})$,
$$
U_{{\bf k} {\bf k}^\prime} ({\bf q}) =
 U_{{\bf k} {\bf
k}^\prime}^{reg} ({\bf q}) + \frac{F ({\bf k}, {\bf k}^\prime, {\bf
q})}{-i \omega + D (\omega, {\bf k} + {\bf k}^\prime)({\bf k} + {\bf
k}^\prime)^2}  \,,
\eqno(9)
$$
entering the Bethe--Salpiter equation and playing the
role of the scattering probability  $W_{{\bf k}
{\bf k}^\prime}$ in the quantum kinetic equation. Neglecting
the spatial dispersion of the diffusion
coefficient\,\footnote{\,Such possibility was
justified in \cite{37}. Attempts to relate the spatial dispersion
of the diffusion coefficient with multifractality of wave
functions \cite{205} ignore the complex-valuedness of the
diffusion coefficient and its complicated rearrangement
near transition \cite{206}. }
and using the estimate in the spirit
of $\tau$-approximation, $D \propto
\langle U \rangle ^{-1}$, where   $\langle ... \rangle$
is  averaging over momenta,
we come to the
self-consistency equation of
the Vollhardt and W${\rm {\ddot o}}$lfle theory
$$
D \sim \left[ U_0 + F_0 \int
\frac{d^dq}{-i \omega + D(\omega, q)q^2} \right]^{-1}  \,.
\eqno(10)
$$
It can be obtained  by approximate solution of
the Bethe--Salpiter equation  \cite{22},
or by the accurate analysis of spectral
properties of the quantum collision operator \cite{37}.
%
It can be written in the physically clear form, if
coefficients are estimated for weak disorder (which is actual for
lower dimensions) and a situation near the band center
in the Anderson model is implied:
$$
\frac{E^2}{W^2} = \frac{D}{D_{min}} + \Lambda^{2-d}
\int\limits_{|q|<\Lambda} \frac{d^dq}{(2\pi)^d}
\,\frac{1}{(-i\omega/D) + q^2}   \,.
\eqno(11)
$$
Here $E$ is the energy of the bandwidth order,  $W$ is the
amplitude of disorder, $\Lambda$ is the ultraviolet cut-off,
$D_{min}$  is a characteristic scale of the diffusion
coefficient corresponding to the Mott minimal
conductivity. Generally, some monotonic function of  $W$
is appearing  in the left-hand side, but it is not
essential for subsequent considerations.

Let introduce the basic integral
$$
I(m) = \int\limits_{|q|<\Lambda} \frac{d^dq}{(2\pi)^d}
\,\frac{1}{m^2 + q^2}   \,,
\eqno(12)
$$
which can be estimated for  $m\ll \Lambda$ as
$$
I(m)=
\left \{ \begin{array}{cc}
c_d/m^{2-d}\,,& d<2 \\
c_2\ln (\Lambda/m)\,,& d=2 \\
I(0)- c_d m^{d-2}\,,& 2<d<4 \\
I(0)- c_4 m^{2}\ln (\Lambda/m)\,,& d=4 \\
I(0)- c_d m^{2}\Lambda^{d-4}\,,& d>4  \end{array} \right.
\,, \eqno(13)
$$
where
$$
c_d=
\left \{ \begin{array}{cc}
\pi K_d/(2\sin(\pi d/2))\,,& d<2 \\
1/2\pi\,,& d=2 \\
\pi K_d/|2\sin(\pi d/2)|\,,& 2<d<4 \\
1/(8\pi^2)\,,& d=4 \\
K_d/(d-4)\,,& d>4  \end{array} \right.
\,, \eqno(14)
$$
and  $K_d=\left[2^{d-1}\pi^{d/2}\Gamma(d/2) \right]^{-1}$ is
the surface  of the $d$-dimensional unit sphere divided by
$(2\pi)^d$.
The metallic phase is possible, when value of $I(0)$ is finite,
i.\,e. for  $d>2$. Accepting $D=const>0$ for $\omega\to 0$
and specifying $\tau$ as a distance to transition, one has
$$
D=D_{min}\, \tau\,,\qquad
\tau = \frac{E^2}{W^2} -I(0) \Lambda^{2-d} \,,
 \eqno(15)
 $$
i.\,e. the exponent of conductivity is unity,  in agreement
with (2). In the dielectric phase we make substitution
$$
D= -i\omega \xi^2\,,\qquad
\xi=m^{-1}  \,,
 \eqno(16)
$$
where $\xi$ is the correlation length. Then Eq.\,11 gives
$$
\xi\sim a \frac{E^2}{W^2}\,,\qquad d=1  \,,
$$
$$
\xi\sim a \exp\left(2\pi\frac{E^2}{W^2}\right)\,,
\qquad d=2  \,,
\eqno(17)
$$
$$
\xi\sim a |\tau|^{-\nu}\,,\qquad d>2     \,,
$$
with the exponent $\nu$ defined by Eq.\,2. In what follows,
we accept $a=\Lambda^{-1}$, so $a$ is  the atomic length
scale, not necessary coinciding with the lattice spacing.

\vspace{4mm}
\begin{center}
{\bf 3. SCALING FUNCTIONS FOR  $D<4$}
\end{center}

\begin{center}
{\bf 3.1. Definition of scaling functions}
\end{center}

For description of quasi-1D systems it is sufficient to
present the basic integral (12) in the following form
$$
I(m) = \frac{1}{L^{d-1}} \sum\limits_{|q_{\bot}|<\Lambda}
\,\int\limits_{-\Lambda}^{\Lambda}\, \frac{dq_{||}}{2\pi}
\,\frac{1}{m^2 + q_{||}^2+ q_{\bot}^2}  \,,\qquad
m^{-1}=\xi_{1D}
\eqno(18)
$$
where  $d$-dimensional vector $q=(q_1,q_2,\ldots,q_d)$ is
replaced by its transversal and longitudinal components
$$
q_{\bot}=(q_1,q_2,\ldots,q_{d-1})\,,
\qquad  q_{||}=q_d\,,
\eqno(19)
$$
and the first is considered as discrete, running the usual
allowed values. The term with  $q_{\bot}=0$ has divergency
$m^{-1}$ for  $m\to 0$, so the system is always localized.

After integration over $q_{||}$,  the following
decomposition is convenient:
$$
I(m) = \frac{1}{ L^{d-1}} \,\frac{1}{\pi m}
\, {\rm arctg} \frac{\Lambda}{ m}\, +
$$
$$
+\frac{1}{\pi L^{d-1}} \sum\limits_{
\begin{array}{c} { \scriptstyle  q_{\bot}\ne 0 } \\
{\scriptstyle |q_{\bot}|<\Lambda  } \end{array}
 }
\left( \,\frac{1}{\sqrt{m^2 + q_{\bot}^2}}
   \, {\rm arctg} \frac{\Lambda}{\sqrt{m^2 + q_{\bot}^2}}\,
-\,\frac{1}{ |q_{\bot}|} \,
    {\rm arctg} \frac{\Lambda}{|q_{\bot}|}
\right)\, +
$$
$$
+\,\frac{1}{\pi L^{d-1}} \sum\limits_{
\begin{array}{c} { \scriptstyle  q_{\bot}\ne 0 } \\
{\scriptstyle |q_{\bot}|<\Lambda  } \end{array}
}
 \frac{1}{ |q_{\bot}|} \,
{\rm arctg} \frac{\Lambda}{| q_{\bot}|} \,\equiv \,
I_1(m)+I_2(m)+I_3(0)
\,.
\eqno(20)
$$
We separated the term with  $q_{\bot}=0$, while  the
remaining sum was rearranged by subtraction and  addition
of the analogous sum with $m=0$.  In the first term one
has trivially
$$
I_1(m) = \frac{1}{ L^{d-2}}
 \left\{ \frac{1}{2 m L} +O\left(\frac{a}{L} \right)
  \right\} \,.
\eqno(20)
$$
The second term can be transformed by taking the limit
$\Lambda\to\infty$ and substituting
$q_{\bot}=2\pi \vec s/L$, where  $\vec s=(s_1,\ldots,s_{d-1})$
is a vector with integer components $s_i=0,\pm 1, \pm 2,
\ldots$:
$$
I_2(m) = \frac{1}{ L^{d-2}} H_0(mL)
 +O\left(m^2 \Lambda^{d-4} \right)   \,,
$$
$$
H_0(z)= \frac{1}{4 \pi} \sum\limits_{ \vec s \ne 0 }
\left( \,\frac{1}{\sqrt{|\vec s|^2 + (z/2\pi)^2}}
-\,\frac{1}{ |\vec s|} \,
\right)\,.
\eqno(21)
$$
The third term can be estimated at $L\to\infty$ by the
replacement of  summation by integration. For finite
$L$ and $d>2$ it has a structure
$$
I_3(0) = \Lambda^{d-2}  \left\{ b_0
+ b_1 \left(\frac{a}{L} \right)^{d-2}
+ b_2 \left(\frac{a}{L} \right)^{d-1}
+\ldots \right\}  \,.
\eqno(22)
$$
Substituting (20--22) in the self-consistency equation
(11), we have for $d>2$
$$
\left(\frac{L}{a} \right)^{d-2}
\left[\tau + O(m^2 a^2) \right]
+ O \left(\frac{a}{L} \right)
= b_1 + H_0(mL) +\frac{1}{2mL}
\eqno(23)
$$
where we replaced
$$
\tau = \frac{E^2}{W^2} - b_0
\eqno(24)
$$
in agreement with definition  (15), since  $b_0$ corresponds
to  $I(0)$, calculated in the integral
approximation. Expressing  $\tau$ through the correlation
length  $\xi$ of the  $d$-dimensional system
($\xi^{-1/\nu} \sim |\tau|=\pm \tau$) and omitting
the terms dissapearing at  $a\to 0$, we have
$$
\pm c_d \left(\frac{L}{\xi} \right)^{d-2}
= H \left(\frac{L}{\xi_{1D}} \right)
\eqno(25)
$$
$$
H(z)=b_1 +\frac{1}{4 \pi} \sum\limits_{ \vec s \ne 0 }
\left( \,\frac{1}{\sqrt{|\vec s|^2 + (z/2\pi)^2}}
-\,\frac{1}{ |\vec s|} \,
\right)\,+\,\frac{1}{2z}
\eqno(26)
$$
i.\,e.  the scaling relation (4) between variables
$\xi_{1D}/L$ and $\xi/L$,  consisting of two branches.

For $d=2$ we have instead (22)
$$
I_3(0) = = \frac{1}{2\pi} \ln\frac{L}{a} + b_1 +\ldots
\eqno(22')
$$
and, using the result from Sec.\,2
$$
\frac{E^2}{W^2} = \frac{1}{2\pi} \ln\frac{\xi}{a} \,\,,
$$
obtain the scaling relation in the form
$$
\frac{1}{2\pi}\ln\left(\frac{\xi}{L} \right)
= H \left(\frac{L}{\xi_{1D}} \right)
\eqno(27)
$$
with the previous definition of $H(z)$. The functions
 $H(z)$ for $d=2$ and $d=3$ are presented
 in Fig.\,2, where   $b_1=0$ was accepted.
\begin{figure}
\centerline{\includegraphics[width=5.1 in]{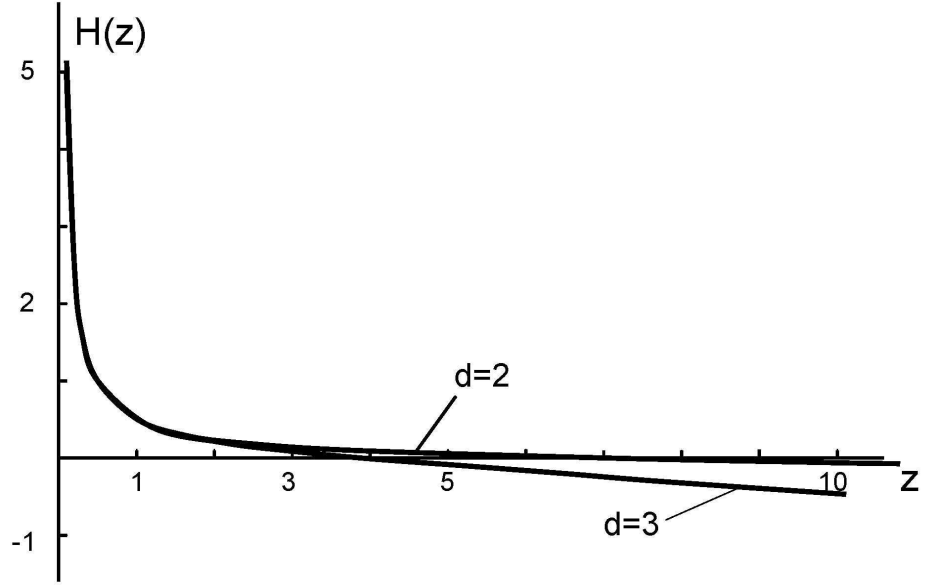}}
\caption{Function $H(z)$ for $b_1=0$ in two and
three dimensions.}
\label{fig2}
\end{figure}

\vspace{2mm}
\begin{center}
{\bf 3.2. Two-dimensional case}
\end{center}

For $d=2$, the constant $b_1$ can be eliminated by the change
of the scale for  $\xi$  (see below) and we can  accept  $b_1=0$.
The asymptotics of $H(z)$ for
$z\ll 1$ is determined by the last term in Eq.\,26,
while for  $z\gg 1$ the sum in Eq.\,26 can be replaced by the
integral
$$
H(z)=
\left \{ \begin{array}{cc}
\displaystyle{\frac{1}{2z}}\,,& \qquad z\ll 1 \\
- \displaystyle{\frac{1}{2\pi}} \ln z + const\,,&
\qquad z\gg 1
\end{array} \right.  \,,
\eqno(28)
$$
so we have in variables  $y=\xi_{1D}/L$ and
$x=\xi/L$
$$
y=\left \{ \begin{array}{cc}
(1/\pi)\ln x \,,& \qquad x\gg 1 \\
const\cdot x\,,& \qquad x\ll 1
\end{array} \right.  \,.
\eqno(29)
$$
The relation between  $x$ and  $y$ for their arbitrary values
can be found by the numerical calculation of the sum in  (26).

The definition of  $\xi_{1D}$ and  $\xi$ in the
Vollhardt and W${\rm {\ddot o}}$lfle theory does not
coincide with one used in numerical experiments. In the former
case,  $\xi^2$ (and analogously  $\xi^2_{1D}$) is defined
as an average $\langle r^2\rangle$  for the localized
eigenfunction $\psi(r)$ \cite{37}.  In the latter case, one has
in mind the definition through the asymptotic behavior
$\exp\{-r/\xi\}$ of the correlation functions, since  $\xi_{1D}$
is calculated as inverse to the minimal Lyapunov
exponent\,\footnote{\,In general, correspondence between
$\xi_{1D}$ and the minimal Lyapunov exponent is not so
straightforward \cite{102}; in the present paper we ignore such
complications.};
the scale of $\xi$ in numerical experiments is
arbitrary from the very beginning. Therefore, in comparison of
theory with numerical results  the scales of
$\xi_{1D}$ and $\xi$ should be chosen from the
best agreement; in the log-log coordinates, such fitting
reduces to parallel shifts along two axes. The general
form of the scaling curve  is determined without adjustable
parameters.
\begin{figure}
\centerline{\includegraphics[width=6.0 in]{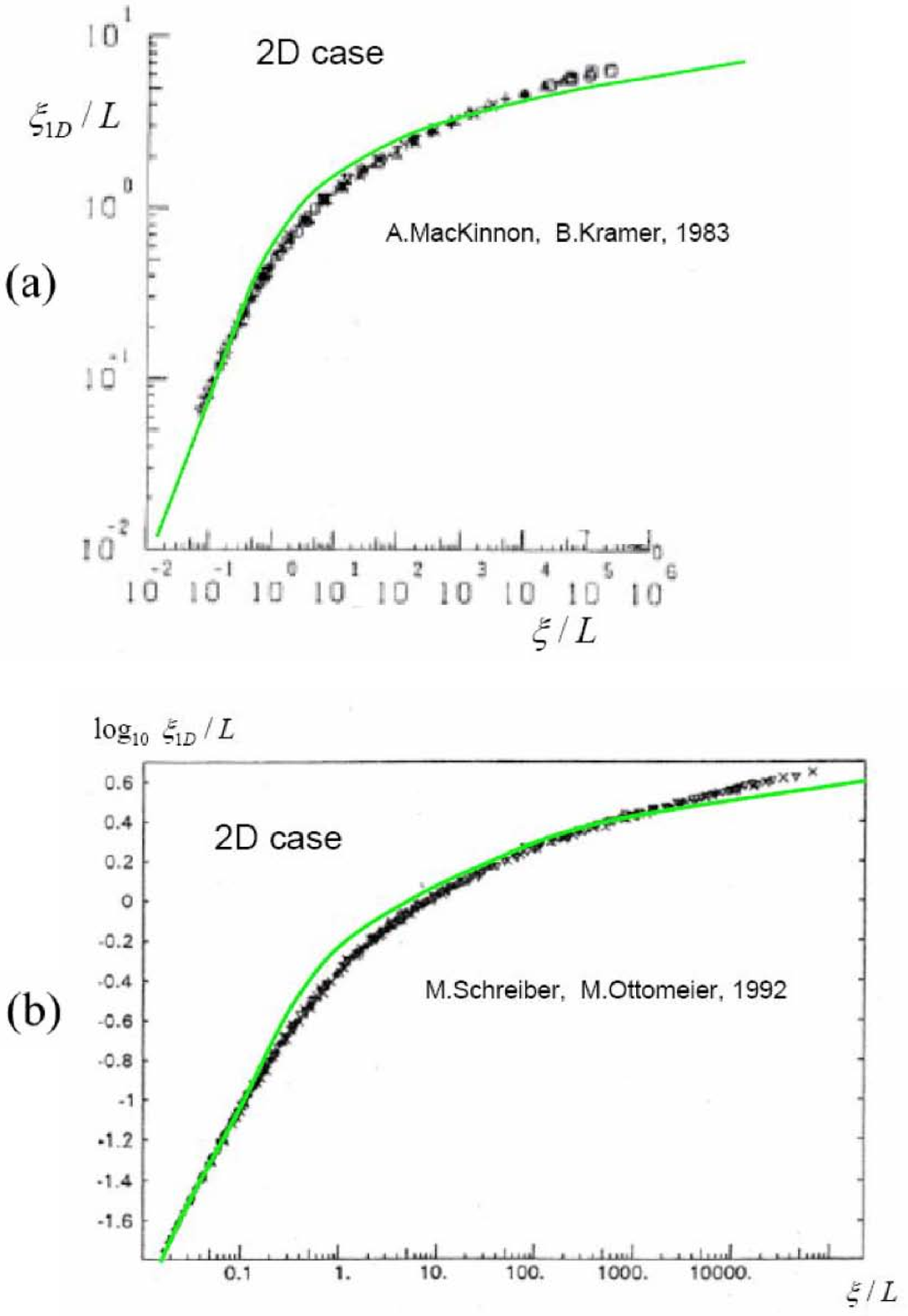}}
\caption{Comparison of the theoretical
scaling curve for $d=2$ with numerical results by
MacKinnon\,--\,Kramer \cite[Fig.\,2,a]{9}  (a) and
Schreiber\,--\,Ottomeier
\cite[Fig.\,4]{10} (b).} \label{fig3}
\end{figure}

In Fig.\,3, the calculated dependence of  $\xi_{1D}/L$ on
$\xi/L$  is compared with the pioneer results by MacKinnon and
Kramer  \cite{9} and the subsequent paper by Schreiber and
Ottomeier \cite{10}, which is cited as the most detailed
investigation of the 2D systems in the framework of the
given algorithm.

\vspace{2mm}
\begin{center}
{\bf 3.3. Three-dimensional case}
\end{center}

The given definition of the sum  $I_3(0)$
implies the choice of cut-off in the form of the
cilindrical domain  ($|q_{\bot}|<\Lambda$, $
|q_{||}|<\Lambda$).  It can be also defined for the
spherical ($|q|<\Lambda$) and cubical ($|q_i|<\Lambda$)
regions:
$$
I_3^{(cub)}(0)
=\,\frac{1}{2\pi^2 L^{d-2}} \sum\limits_{
\begin{array}{c} { \scriptstyle  \vec s\ne 0 } \\
{\scriptstyle |s_i|<\Lambda L/2\pi } \end{array}
}
 \frac{1}{ |\vec s|} \,
{\rm arctg} \left( \frac{\Lambda L}{2\pi |\vec s|}
 \right)
$$
$$
I_3^{(cil)}(0)
=\,\frac{1}{2\pi^2 L^{d-2}} \sum\limits_{
\begin{array}{c} { \scriptstyle  \vec s\ne 0 } \\
{\scriptstyle |\vec s|<\Lambda L/2\pi } \end{array}
}
 \frac{1}{ |\vec s|} \,
{\rm arctg} \left( \frac{\Lambda L}{2\pi |\vec s|}
 \right)
\eqno(30)
$$
$$
I_3^{(sph)}(0)
=\,\frac{1}{2\pi^2 L^{d-2}} \sum\limits_{
\begin{array}{c} { \scriptstyle  \vec s\ne 0 } \\
{\scriptstyle |\vec s|<\Lambda L/2\pi } \end{array}
}
 \frac{1}{ |\vec s|} \,
{\rm arctg} \left(
\frac{\sqrt{(\Lambda L/2\pi)^2-|\vec s|^2}}
{|\vec s|} \right)
$$
Numerically we have for these three cases
$$
I_3(0)=\left \{ \begin{array}{cc}
\displaystyle{ 0.0618 \Lambda - 0.180 L^{-1}}
\,& \quad (cube) \\
\displaystyle{ 0.0573 \Lambda - 0.314 L^{-1}}
\,& \quad (cilinder) \\
\displaystyle{ 0.0507 \Lambda - 0.310 L^{-1}}
\,& \quad (sphere)\end{array} \right.  \,,
\eqno(31)
$$
i.\,e. value of $b_1$ is not universal but depends on the way of
cut-off. The change of this constant allows to make the scaling
curve more symmetric, or less symmetric; it was chosen
from the best agreement\,\footnote{\,The fitting was
made by hand (using several
reference points) and probably is not optimal.},
though its variation in the interval
$(-0.3,\, 0)$ does not affect the results significantly.
As in the $2D$ case, the absolute scales for  $\xi$ and
$\xi_{1D}$ are not fixed by the theory.

Using the asymptotic behavior of  $H(z)$
$$
H(z)=
\left \{ \begin{array}{cc}
\displaystyle{1/2z}\,,& \qquad z\ll 1 \\
-A(z-z^*)\,,& \qquad z\to z^* \\
-c_d z^{d-2}\,,& \qquad z\gg 1
\end{array} \right.  \,,
\eqno(32)
$$
one has in variables  $y=\xi_{1D}/L$ and $x=\xi/L$
$$
y=\left \{ \begin{array}{cc}
2c_d/x^{d-2} \,,& \qquad y\gg 1 \\
y^*\pm B/x^{d-2} \,,& \qquad y\to y^* \\
x \,,& \qquad y\ll 1 \\
\end{array} \right.  \,,
\eqno(33)
$$
where $z^*$ and $y^*=1/z^*$ are  values of $z$ and  $y$ in the
critical point. The same relation, considered in variables $y$
and $1/x$, determines the $L$-dependence of the scaling parameter
(Fig.\,1)  giving two universal curves for $\tau>0$ and
$\tau<0$  to which all other curves are reduced by the
the scale transformation:
$$
y=\frac{\xi_{1D}}{L}=
\left \{ \begin{array}{cc}
\sim \tau L^{d-2} \,,& \qquad y\gg 1 \\
y^* + const\cdot \tau L^{d-2} \,,& \qquad y\to y^*\\
\xi/L \,,& \qquad y\ll 1
\end{array} \right.  \,.
\eqno(34)
$$

\begin{figure}
\centerline{\includegraphics[width=5.8 in]{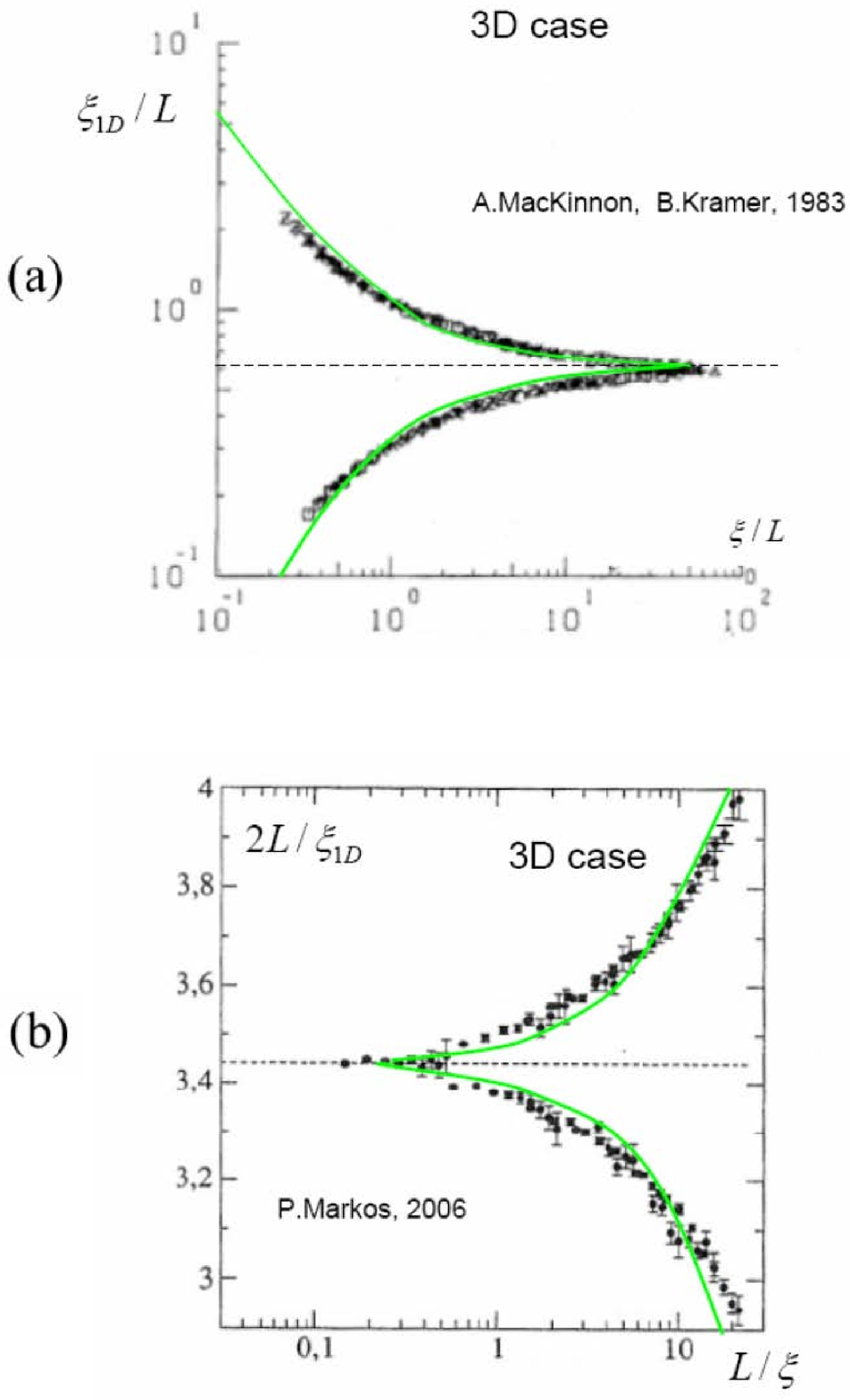}}
\caption{Comparison of the theoretical
scaling curves for $d=3$ with numerical results by
MacKinnon\,--\,Kramer \cite[Fig.\,2,b]{9}  (a) and
Markos \cite[Fig.\,53,\,right]{2}
(b). The values $b_1=-0.240$ and  $b_1=-0.0718$ were used in the
former and the latter case correspondingly.} \label{fig4}
\end{figure}
In Fig.\,4, the obtained scaling curves are  compared with the
early  results by MacKinnon -- Kramer  \cite{9} and the
more precise results by Markos  \cite{2}. In the former case the
agreement is satisfactory, in the latter case there is
discrepancy on the level of 2\,--\,3 standard deviations. However,
one should have in mind how scaling curves are constructed:
the $L$-dependences for different   $\tau$ are
"measured" in the interval $(L_{min},L_{max})$, and then they
are fitted to each other by a change of the scale (Fig.\,5).
\begin{figure}
\centerline{\includegraphics[width=5.1 in]{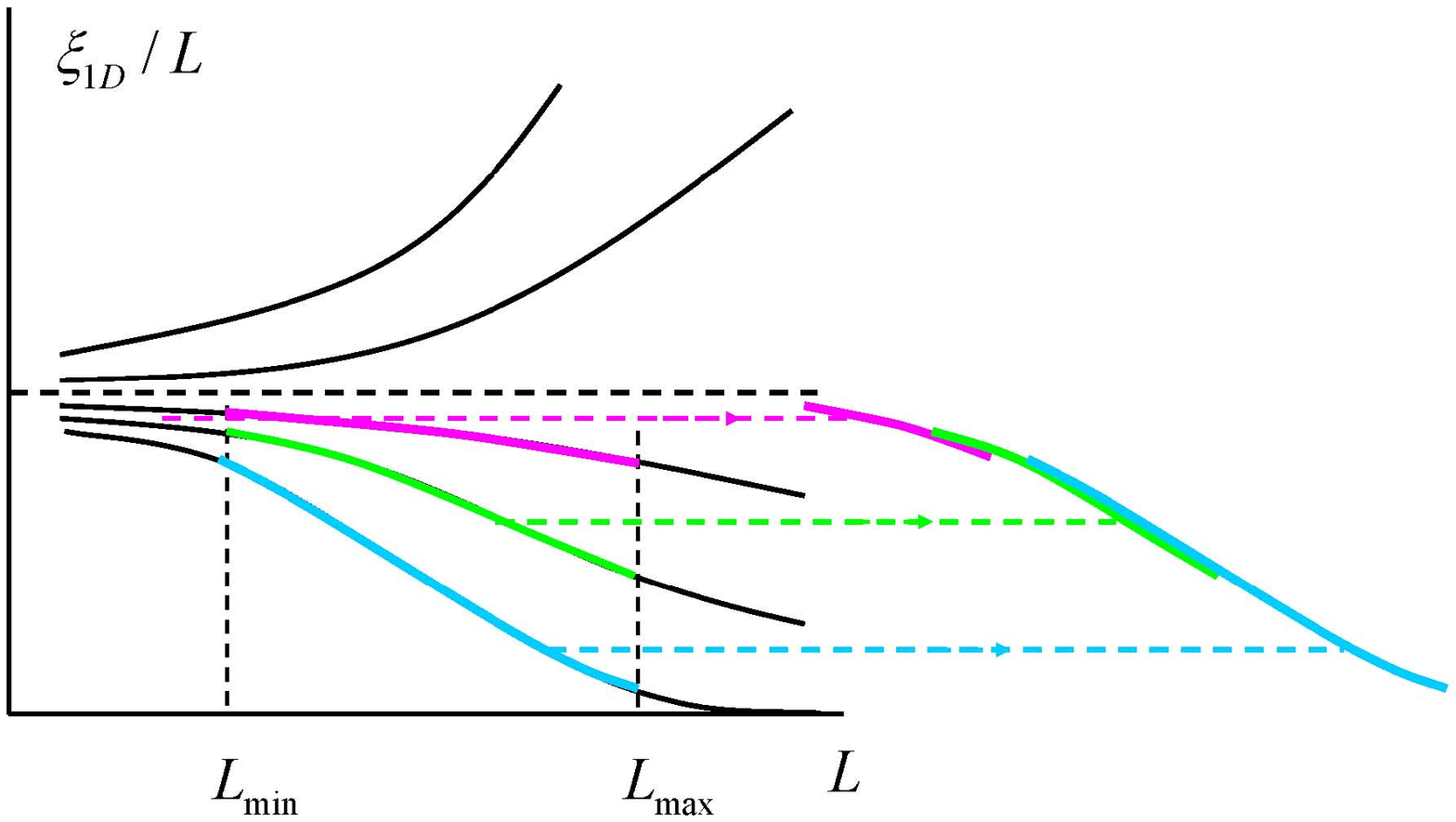}}
\caption{Construction of scaling curves.}
\label{fig5}
\end{figure}
The
full scaling curve is never present in  one experiment, and only
separate fragments of it are measured. It is clear from Fig.\,4,b
that the change of the scale along the horizontal axis
(reducing to a parallel shift in the logarithmic coordinates)
allows to obtain  satisfactory fits for the left, right or
middle portion of the curve. It looks, there are no serious
contradictions with the  Vollhardt and W${\rm {\ddot o}}$lfle
theory on the level of raw data.

\vspace{4mm}

\begin{center}
{\bf 4. DISCUSSION OF THE SITUATION AT $D=3$}
\end{center}

The interesting question arises: if the  Vollhardt and
W${\rm {\ddot o}}$lfle theory  describes the raw
data successfully, then why all numerical experiments give
$\nu>1$ for $d=3$ ?

The history of this question goes back to two papers  \cite{8}
and \cite{9} by MacKinnon and Kramer, based on the same array
of the data. The first of them gives the result
$$
\nu=1.2\pm 0.3 \,,
\eqno(35)
$$
compatible with the value $\nu=1$; the second paper confirms
this result for a certain fitting procedure, but reports
the "more precise" result
$$
\nu=1.50\pm 0.05  \,,
\eqno(36)
$$
corresponding to the most extremal of present-day values.
The first result is based on the analysis of the scaling
curve, whose compatibility  with the  Vollhardt and W${\rm
{\ddot o}}$lfle theory  is clear from Fig.\,4,a  and
is confirmed by the authors themselves. Further, they
indicate that scaling is not satisfactory in the small vicinity
of the critical point, and this vicinity was discarded in
their treatment.  In fact, such situation is  natural, because
the small vicinity of the transition is strongly affected by
scaling corrections (see Eq.\,23); the latter are small
in magnitude but should
be compared with the small value of  $\tau$.  However, the
authors of \cite{8,9} estimated this situation as internally
inconsistent and suggested another treatment procedure,
which is specially  based on the analysis of that small region
where scaling is absent. Already at this stage it  is possible to
understand that the latter procedure is not satisfactory.

Indeed, using the systems of the restricted size  $L$, one
can work  straightforwardly  only in the regime
$\xi\alt L$, since in another case the correlation functions are
strongly affected by finiteness of the system.  The use of
finite-size scaling allows "to jump above the head" and to
advance in the region  $\xi\agt L$; however, it is possible only
if (a) scaling exists theoretically, and (b) it is observed
empirically. If any of two conditions is violated, no such
advancement is possible, and one is unable to obtain any
experimental information on the large  $\xi$ region;
any manipulations in this region become irrelevant. This
conclusion is valid in respect of the result (36), since absence
of scaling is admitted by the authors. The same conclusion
follows from the common sense: if value $\nu=1$ is compatible
with the scaling curve, then all the more it is compatible with
the raw data (see the end of Sec.\,3). However, this value is
rejected by the result (36), and hence the latter  should
be qualified as essentially incorrect.

The indicated tendencies was continued in other papers. The
treatment based on the scaling curves gave rather conservative
estimates\,\footnote{\,As
was discussed in  \cite{14}, the estimate of
$\nu$ depends on the fragment of the scaling curve, which
is used for fitting.},
not very different from (35).
The results close to (36) were stabilized only when
the control of scaling ceased to be imperative
and the analysis of small vicinity of the critical point was
generally accepted.

The latter procedure is based on
representation of (4) in the form
$$
\frac{\xi_{1D}}{L} =
F\left(\frac{L^{1/\nu}}{\xi^{1/\nu}} \right) =
F\left( \tau L^{1/\nu} \right) \approx
y^* +A \tau L^{1/\nu}+\ldots \,
\eqno(37)
$$
i.\,e.  the regular expansion in  $\tau$ is used,
motivated by the absence of phase transitions in quasi-1D
systems; then the derivative over  $\tau$ behaves as $L^{1/\nu}$
and gives the exponent $\nu$ straightforwardly.
Such treatment is correct if the scaling relation (4) is exact.
However, it is not exact: linearization of  (23) gives
$$
\frac{\xi_{1D}}{L} = y^* +
A\left(\frac{L}{a} \right)^{d-2}
\left[\tau -c\frac{a^2}{\xi_{1D}^{2}}  \right]
+ O \left(\frac{a}{L} \right) \,.
\eqno(38)
$$
Differentating  over  $\tau$ and excluding
$(\xi_{1D})'_\tau$ from the right-hand side
in the iterative manner,
one has
$$
\left(\frac{\xi_{1D}}{L}\right)'_\tau = A_0 L^{d-2}
+ A_1 L^{d-6}   \,.
\eqno(39)
$$
Producing subsequent iterations and taking into account
further corrections to scaling, one have the following
structure of the result
$$
\frac{\xi_{1D}}{L}-y^*= \tau \left\{ A_0 L^{1/\nu}
+ A_1 L^{\omega_1} + A_2 L^{\omega_2} +\ldots\right\}
+B_1 L^{-y_1} + B_2 L^{-y_2} +\ldots \,,
\eqno(40)
$$
which can be obtained from the general considerations
based on the Wilson renormalization group \cite{21}.

In three dimensions, the main scaling correction in (39)
reduces to constant,
and hence
$$
\frac{\xi_{1D}}{L}-y^*= A \tau \left( L + L_0\right)
 \,,
 \eqno(41)
$$
where the terms dissapearing at $L\to\infty$ are neglected.
It is clear from Fig.\,6,a, that numerical data by Markos
\cite{2} are excellently fitted by (41). The author
himself interpreted them in accordance with (37)
and also had the good fitting (Fig.\,6,b).
\begin{figure}
\centerline{\includegraphics[width=6.0 in]{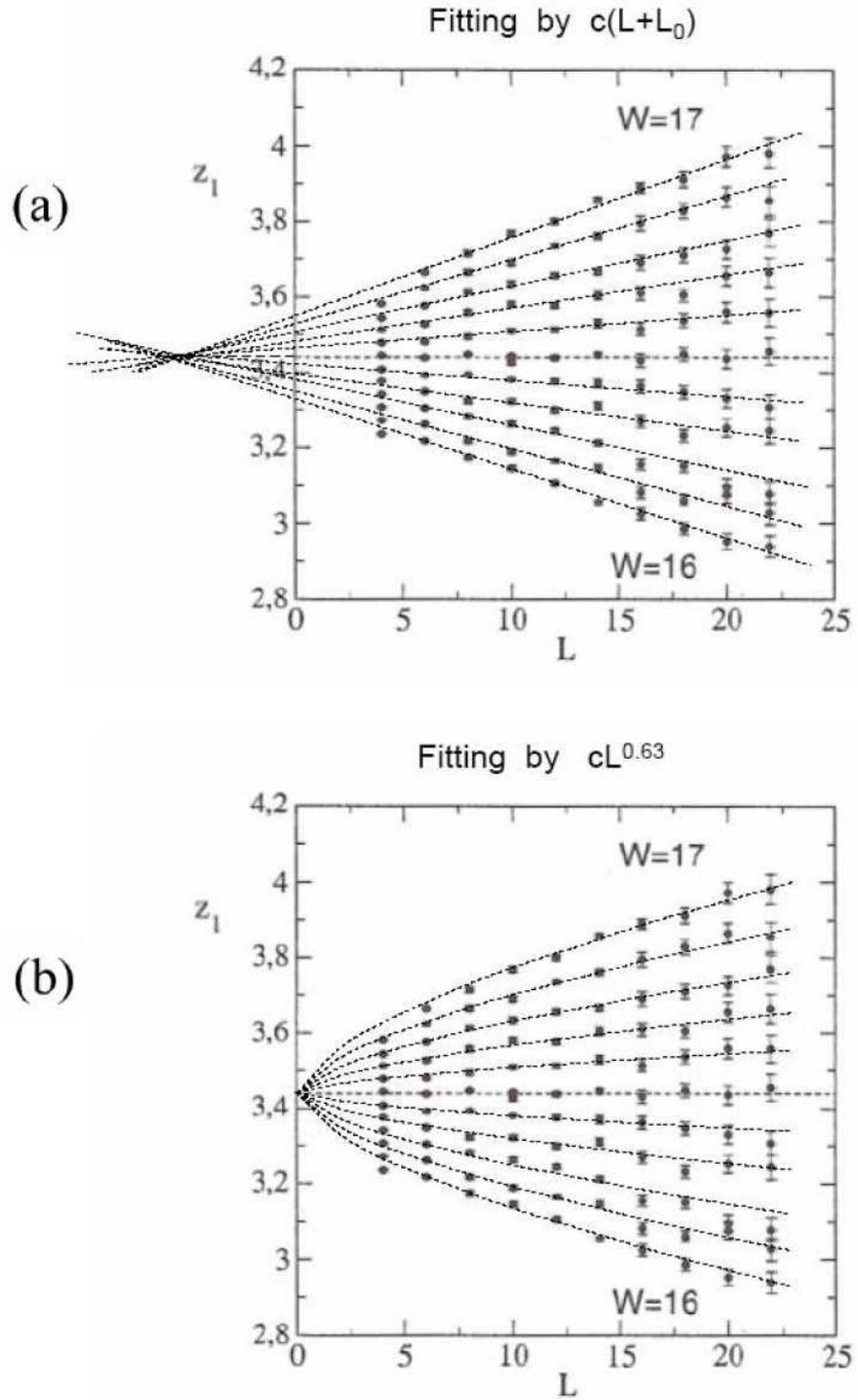}}
\caption{Numerical data by Markos for  $z_1=2L/\xi_{1D}$
in the small vicinity of the critical point
 \cite[Fig.\,53,\,left]{2} and their fitting by dependences
  $'(L+L_0)$ (a) and  $'L^{0.63}$  (b).} \label{fig6}
\end{figure}

Such ambiguity of interpretation
has a general character. If the
combination  $A_1 L^{\beta_1}+A_2 L^{\beta_2}$
can be linearized  in the log-log coordinates
with the average slope $(\beta_1+\beta_2)/2$
and the accuracy $\epsilon$, then variation  $\beta_1\to
\beta_1+\delta$, $\beta_2\to \beta_2-\delta$ preserves
linearity on the same level of accuracy, till  $|\delta|\alt
|\beta_1-\beta_2|/2$.
 If several terms are
retained
 in  (40),
the situation becomes
not controllable at all: non-linear
fitting with minimization of $\chi^2$ reveals the
huge number of minima, and the deepest of them is not
necessary correct; a vicinity of any minimum is acceptable
if it is satisfied to the  $\chi^2$ criterion.
Analysis of all such minima is impossible, and there is no honest
procedure to deal with such situation\,\footnote{\,These
questions were
discussed  \cite{21}  in relation with the paper
\cite{19}. Nevertheless, this paper is continued to be cited
 \cite{2} as a prominent achivement.}. In conclusion, the
 conventional treatment is heavily based on the assumption, that
 only the main term in (40) is essential; the problem of
 fitting becomes hopeless, if additional terms are not
 negligible.

In the framework of  the Vollhardt and W${\rm {\ddot o}}$lfle
theory, we have a completely consistent picture. The
quantity $L_0$ violates scaling and empirically has rather
large value, $L_0\approx 5 $ (in lattice units). The good scaling
is possible only for $L\gg 5$, and even for the largest systems
($L=20\div 30 $) deviations of scaling are described by the
parameter $L_0/L\sim 0.2$; so  discrepancies in Fig.\,4,b should
not be of any surprise.  The theoretical value of  $L_0$ is of
the order $\Lambda^{-1}$ with the coefficient depending on the
way of cut-off; it is essential that $L_0$ is positive and
limited from below by the atomic scale.

\vspace{4mm}

\begin{center}
{\bf 5. SCALING FOR HIGHER DIMENSIONS} \end{center}

\begin{center}
{\bf 5.1. Dimensions  $d>4$}
\end{center}

For $d\ge 4$, the sum  $I_2(m)$ is divergent at the upper limit
and the cut-off parameter $\Lambda$ cannot be considered as
infinite.  For the accurate trasformation, we introduce the scale
$\Lambda_1$ such as
$$
m\ll \Lambda_1\ll \Lambda
\eqno(42)
$$
and divide summation in $I_2(m)$ into two regions
$|q_{\bot}|<\Lambda_1$ and $|q_{\bot}|>\Lambda_1$. In the first
region we  use
that $|q_{\bot}|\ll \Lambda$, so as
$$
I_2^{(1)}(m) =-m^2 \frac{1}{2 L^{d-1}} \,
 \sum\limits_{
\begin{array}{c} { \scriptstyle  q_{\bot}\ne 0 } \\
{\scriptstyle |q_{\bot}|<\Lambda_1  } \end{array}
 }
 \,\frac{1}{ |q_{\bot}| \, \sqrt{m^2 + q_{\bot}^2}
 \,\left(|q_{\bot}| + \sqrt{m^2 + q_{\bot}^2}\right)}
\eqno(43)
$$
and
$$
I_2^{(1)}(m) = \left\{
\begin{array}{cc}
{ \displaystyle
-m^2 \left\{
\frac{K_{d-1}\Lambda_1^{d-4}}{4(d-4)} +
O \left( m^{d-4} \right)
\right\}}\,,\qquad & m\agt L^{-1} \\
{\displaystyle
-m^2 \left\{
\frac{K_{d-1}\Lambda_1^{d-4}}{4(d-4)} +
O \left( L^{4-d} \right)
\right\}  }\,,\qquad  & m\alt L^{-1}
\end{array}  \right. \,,
\eqno(44)
$$
%
i.\,e. the result is obtained analytically
(in the main approximation)
for the arbitrary relation between  $m$ and $L^{-1}$.
Indeed, for $m\gg L^{-1}$  the sum is estimated by the integral,
which is converging at the lower limit already for $m=0$; so
a finiteness of  $m$ gives only small corrections. In the case
$m\alt L^{-1}$, the main effect from a finiteness of  $L$
is related with the absence of the term  $ q_{\bot}=0$,
which can be estimated as restriction  $ |q_{\bot}|\agt L^{-1}$
in the integral approximation.

In the region  $ |q_{\bot}|>\Lambda_1$ we make use of
condition $ |q_{\bot}|\gg m$  and produce expansions in  $
m/|q_{\bot}|$; after separation
the factor  $m^2$
we can set $m=0$ in the sum and estimate it by transformation
to the integral
$$
I_2^{(2)}(m) = m^2
\frac{K_{d-1}\Lambda_1^{d-4}}{4(d-4)}
 - c m^2 \Lambda^{d-4}
\eqno(44)
$$
where $c$ depends on the way of cut-off; dependence on
$\Lambda_1$ dissapears in the sum $I_2^{(1)}+I_2^{(2)}$.

The results for  $I_1(m)$ and $I_3(0)$ are the same as in
Sec.\,3. The self-consistency equation takes the form
$$
\tau \Lambda^{d-2} =
\frac{1}{L^{d-2}}\, \frac{1}{2 m L} - c m^2 \Lambda^{d-4}
\eqno(45)
$$
Substituting  $\tau \sim \xi^{-2}$ and introducing variables
$$
y=\frac{\xi_{1D}}{L} \left(\frac{a}{L} \right)^{(d-4)/3}\,,\qquad
x=\frac{\xi}{L} \left(\frac{a}{L} \right)^{(d-4)/3}\,,
\eqno(46)
$$
we obtain the scaling relation in the analytical form
$$
\pm \frac{1}{x^2} = y -\frac{1}{y^2}
\eqno(47)
$$
%
where all coefficients are made equil to unity by redefinition
of the  scales for $\xi_{1D}$ and $\xi$.
Relations  (46\,,47) contain the atomic scale
$a$, as was expected from non-renormalizability of theory
(Sec.\,1).

According to (46\,,47), the role of the scaling parameter is
played by the quantity $y$ instead of $\xi_{1D}/L$;
the $L$-dependence  of $y$
is analogous to Fig.\,1, i.\,e. all curves
corresponding to $\tau>0$ and $\tau<0$ can be  reduced
to two universal ones by the scale transformation. The transition
point corresponds to  $y=1$, so as
$$
\frac{\xi_{1D}}{L} \sim \left(\frac{L}{a} \right)^{(d-4)/3}
\,,\qquad \tau=0
\eqno(48)
$$
%
and the critical point cannot be fixed by the condition
$\xi_{1D}/L=const$.

\begin{center}
{\bf 5.2. Four-dimensional case}
\end{center}

In the case $d=4$ we have analogously
$$
I_2(m) = \left\{
\begin{array}{cc}
{ \displaystyle  -c_4\, m^2
\ln\frac{\Lambda}{m} +
O \left( 1 \right)
\,,}\qquad & m\agt L^{-1} \\
{      }\\
{ \displaystyle  - c_4\, m^2
\ln(\Lambda L) +
O \left( 1 \right)
}\,,\qquad
 & m\alt L^{-1}
\end{array}  \right. \,,
\eqno(49)
$$
i.\,e. two results differ by  $\ln(mL)$, which reduces to
the double-logarithmic quantity in the actual region (see below).
Neglecting such quantities, we can obtain scaling also for
$d=4$. The self-consistency equation has a form
$$
\tau \Lambda^{2} =
\frac{1}{2 m L^3} -
c_4\, m^2 \ln\frac{\Lambda}{m}
\eqno(50)
$$
and after
changing to
$\xi$ and $\xi_{1D}$
$$
\pm \frac{c_4}{\xi^2} \ln( \xi/a)=
\frac{\xi_{1D}}{2  L^3} -
\frac{c_4}{\xi_{1D}^2}  \ln(\xi_{1D}/a)
\eqno(51)
$$
The scaling relation (47) is obtained in variables
$$
y=\frac{\xi_{1D}}{ L} \left[\ln(L/a)\right]^{-1/3}
\,,\qquad
x=\frac{\xi}{\left[\ln(\xi/a)\right]^{1/2} }
  \frac{\left[\ln(L/a)\right]^{1/6}}{ L }     \,,
\eqno(52)
$$
i.\,e. the scaling parameter  $y$ is logarithmically modified
in comparison with $\xi_{1D}/L$ and should be considered as
a function of "the modified length"
$\mu(L)=L \left[\ln(L/a)\right]^{-1/6}$; then
all dependences become analogous to Fig.\,1 and a change of
the scale for  $\mu(L)$ allows to reduce them to two
universal curves for  $\tau>0$ and $\tau<0$. The critical
point corresponds to $y=const$, so as
$$
\frac{\xi_{1D}}{L} \sim \left(\ln\frac{L}{a} \right)^{1/3}
\,,\qquad \tau=0
\eqno(53)
$$
i.\,e. parameter $\xi_{1D}/L$ grows logarithmically
in the transition point.

\begin{center}
{\bf 5.3. Modified scaling for $d=4-\epsilon$}
\end{center}

From the methodical point of view, it is interesting to
derive the modified scaling for $d=4-\epsilon$; in this case, the
sum $I_2(m)$  converges  formally at the upper limit, but this
convergence is slow and a finiteness of  $\Lambda$ gives the
essential effect. Analogously to (49) we have
$$
I_2(m) = \left\{
\begin{array}{cc}
{ \displaystyle  -c_4\, m^2
\frac{m^{-\epsilon}-\Lambda^{-\epsilon}}{\epsilon}
\,,}\qquad & m\agt L^{-1} \\
{ \displaystyle  - c_4\, m^2
\frac{L^{\epsilon}-\Lambda^{-\epsilon}}{\epsilon}
}\,,\qquad
 & m\alt L^{-1}
\end{array}  \right. \,,
\eqno(54)
$$
and the scaling relation (47) is obtained in variables
$$
y=\frac{\xi_{1D}}{ L}
\left[\frac{\epsilon}{1-(L/a)^{-\epsilon}}\right]^{1/3}
\,,\qquad
x=  \frac{\epsilon^{1/3}
(\xi/a)}{\left[(\xi/a)^{\epsilon}-1\right]^{1/2} } \,
  \frac{\left[1-(L/a)^{-\epsilon} \right]^{1/6}}
  {   (L/a)^{1-\epsilon/2} } \,.
\eqno(55)
$$
Once  again we have the modified scaling parameter $y$
and "the modified length"
$\mu(L)=L^{1-\epsilon/2}
\left[1-(L/a)^{-\epsilon}\right]^{-1/6}$, in terms of which
the dependences of Fig.\,1 are recovered.  In the critical point
we have  $y=1$  and
$$
\frac{\xi_{1D}}{ L}=
\left[\frac{1-(L/a)^{-\epsilon}}{\epsilon}\right]^{1/3}
\, = \left\{
\begin{array}{cc}
{ \displaystyle
\left[\ln (L/a) \right]^{1/3}
\,,}\qquad & \ln(L/a)\alt 1/\epsilon \\
{    }\\
{ \displaystyle
\left( 1/\epsilon \right)^{1/3}
}\,,\qquad
 & \ln(L/a)\agt 1/\epsilon
\end{array}  \right. \,,
\eqno(56)
$$
\begin{figure}
\centerline{\includegraphics[width=5.0 in]{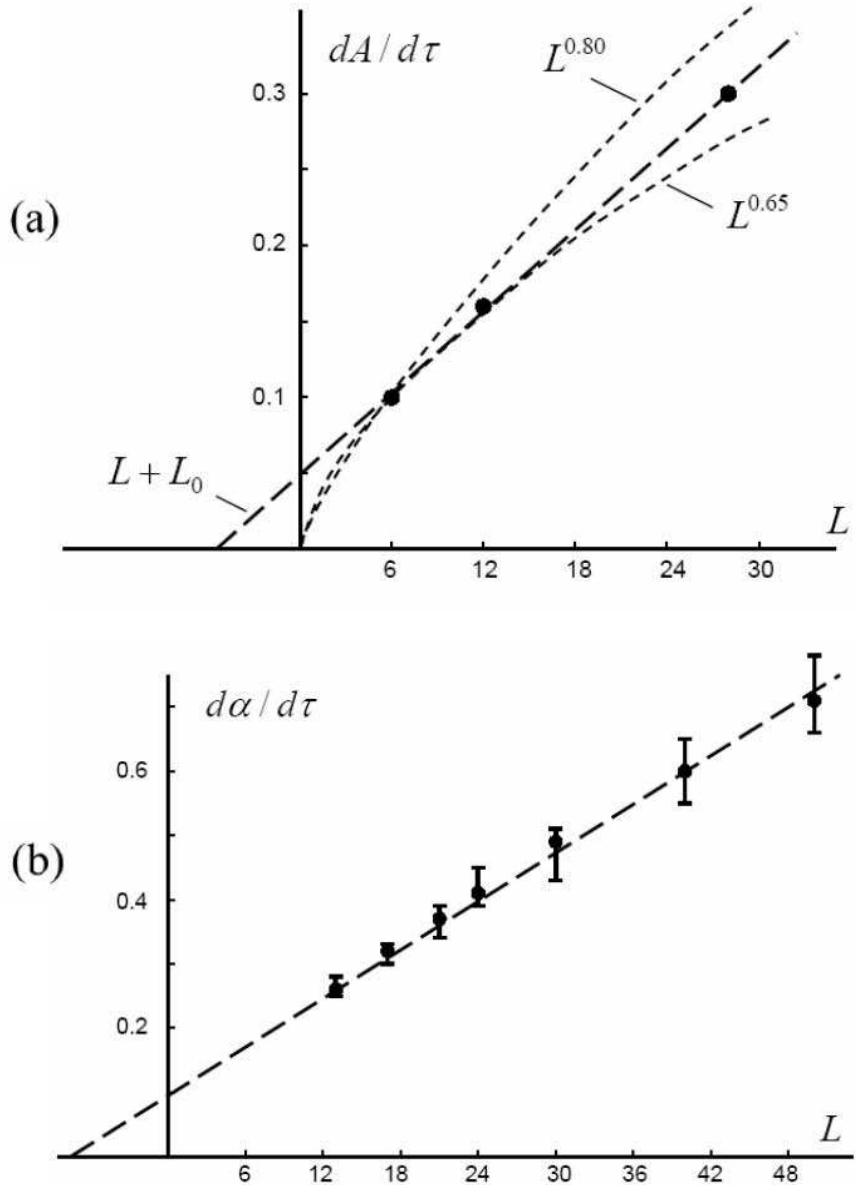}}
\caption{\small Fitting by dependence $'(L+L_0)$
(dashed line) for numerical data, based on
 the level statistics:
 (a) Data by Zharekeshev and Kramer
 \cite{18a}. The points correspond to the average
derivatives of the scaling parameter  $A$ (arbitrary units),
determined from Fig.\,4  of \cite{18a} in the interval $16<W<17$.
A statistical error related with each point can be estimated very
conservatively (see Table in \cite{21}) due to the irregular
character of curves given in the indicated figure; uncertainty
allowed by the authors themselves corresponds to the gap between
dependences  $L^{0.80}$ and $L^{0.65}$, determining the upper and
lower bound of the result for the critical exponent, $\nu=1.40\pm
0.15$. (b) Data obtained by Schreiber' group \cite{207}; the
points correspond to the derivative of the scaling parameter
$\alpha$ (arbitrary units) determined by the slope of solid lines
in the inset of Fig.\,3 in \cite{207}; their uncertainty is
obtained by variation of the slope allowed by the size of
experimental points.} \label{fig7}
\end{figure}
i.\,e. parameter  $\xi_{1D}/L$ grows logarithmically
till the large length scale $L_0\sim a \exp\{const/\epsilon\}$,
and then saturates at the constant value. Such kind of
scaling is useful as an alternative treatment for  $d=3$,
in order to investigate the systematic errors related with
the possible existence of the large length scale. In this case
the parameter  $a$ in fact corresponds to  $L_0$ and can
be essentially different from the lattice constant; it should
be adjusted  from the condition of the best
quality of scaling. There is no need to be
bound by Eq.\,47, which is valid for small $\epsilon$;
it is  more reasonable to determine the relation $y=F(x)$
empirically.  As for expressions  (55), their
extrapolation to $\epsilon=1$ does  not present any problem,
since for $L,\,\,\xi\gg a$ the modified  scaling safely reduces
to the usual one (see  Eq.\,4). In fact, it is identical
to (4) if no large  scale $L_0$ is present.  However, in the
presence of the large length $L_0$ such scaling is more adequate
than (4).

\vspace{4mm}

\begin{center}
{\bf 6. CONCLUSION}
\end{center}

The above analysis allows to conclude that
the Vollhardt--W${\rm {\ddot o}}$lfle theory
does not have essential contradictions with numerical results
on the level of  raw data. The different critical behavior
reported usually in numerical papers originates from the
fact, that some time ago the pure "experimental" approach to the
problem was rejected and replaced by phenomenological
analysis,
which is practically hopeless
in the corresponding region. In particular, dependence  $L+L_0$
with $L_0>0$ is interpreted as  $L^{1/\nu}$ with $\nu>1$.

We have restricted our discussion by the widespread variant of
finite-size scaling, based on application of auxiliary
quasi-1D systems. Apart it, another algorithms are used, based on
the level statistics \cite{12},
the conductance distribution, the mean conductance, etc.
\cite{2}. The scaling curves calculated above are not universal
and cannot be used for comparison with such results.  The scaling
for higher dimensions is also not universal: for example, another
behavior in the critical point is expected for the Thouless
parameter \cite{204}. The only exclusion is the result (40),
which remains unchanged in all cases.
Indeed, this result can be
obtained from the general arguments based on the Wilson
renormalization group \cite{21}; the exponents $\omega_1$,
$\omega_2$, $y_1$, $y_2,\,\ldots$ are determined by scaling
dimensions of irrelevant parameters  and hence are
universal. Correspondingly, the result (41) is unchanged, which
explain the origin of the effective values $\nu>1$ (Fig.\,7).

Fig.7,a can be considered as an etalon illustration,
corresponding to
most  of numerical papers. Indeed, there is an
overall consensus that data for  $L\alt 5$ fall out of the
scaling picture and should be discarded; large systems with  $L
\agt 30$ are practically never used;
the error corridor between
dependences $L^{0.80}$  and  $L^{0.65}$ corresponds to the
typical accuracy of numerical papers. Fig.7,b illustrates one of
the rare papers treating  the systems of the record  size
\cite{207}.  Finally, Fig.8 shows the rare example of
high-precision data \cite{208}.
\begin{figure}
\centerline{\includegraphics[width=7.0 in]{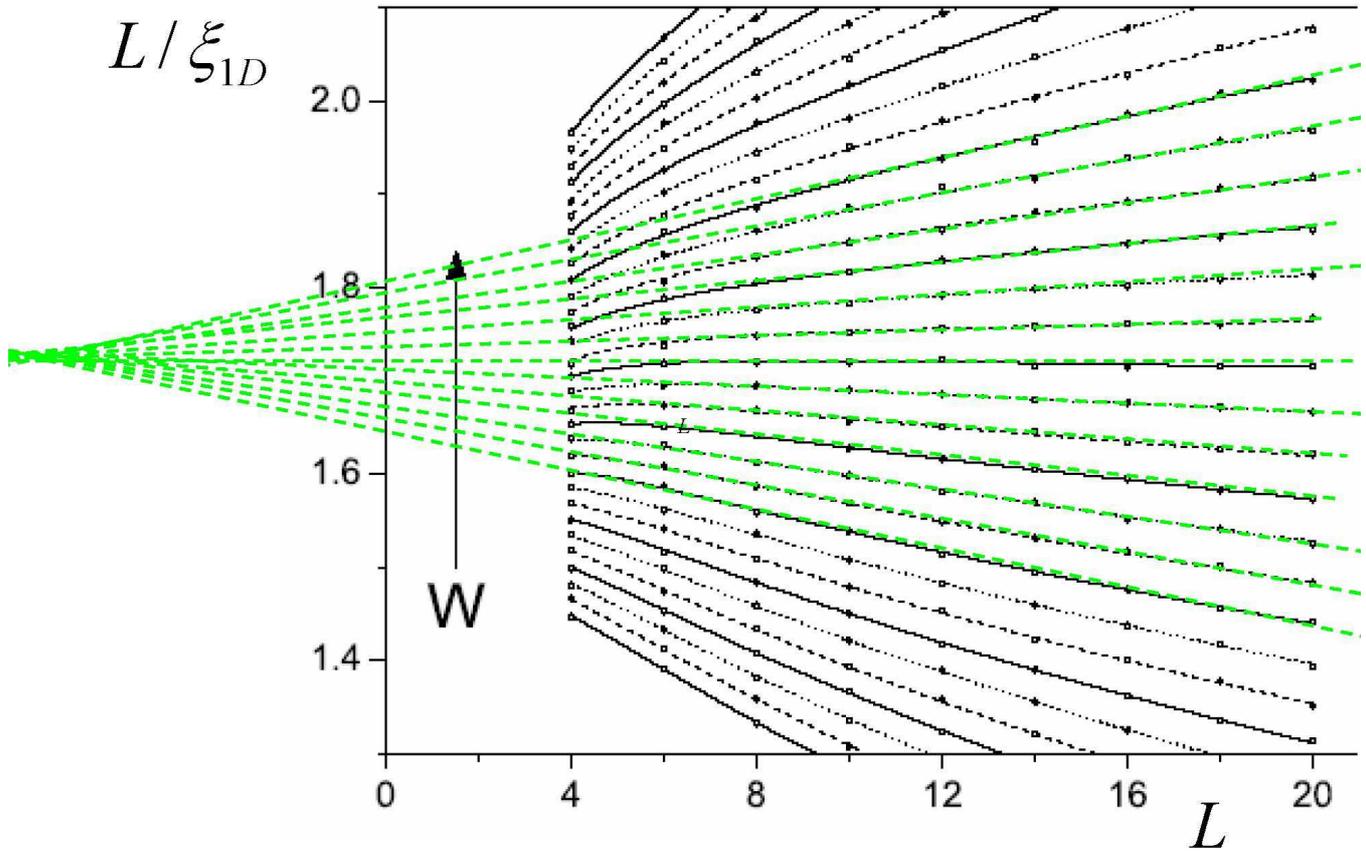}} \caption{
High-precision data by Kramer et al \cite{208} and their fitting
by dependence  $C(L+L_0)$ (green lines). Black solid and dashed
lines correspond to the ambiguous many-parameter treatment  with
$\nu=1.57$. Corrections $O(a/L)$ are visible for small $L$.   }
\label{fig7} \end{figure}

Our final remark is as follows. Even if subsequent investigations
reveal that  the Vollhardt and W${\rm {\ddot o}}$lfle theory
is not
exact, nevertheless no
confidence can be given to the present-day estimates of the
exponent $\nu$ \cite{8}--\cite{20a}. Fig.\,6 clearly demonstrates
that values  $\nu\approx 1.6$ and $\nu=1$ are equally compatible
with the raw data, and hence the treatment procedure is extremely
ambiguous.


\end{document}